# Deterministic patterning and alignment of tellurium quantum wires using nanoscale templates

I K M Reaz Rahman[1,2], Scott Dhuey[3], Moniruzzaman Jamal[2,3,4], Taehoon Kim[1,2], Naoki Higashitarumizu[1,2], Selven Virasawmy[3], Aidar Kemelbay[3], Genki Ohkatsu[1,2,5], Aditya Nirmale[5], Inha Kim[1,2], Hyong Min Kim[1,2], Karen C. Bustillo[3], Joel W. Ager III[2,4], Daryl C. Chrzan[2,4], Mary Scott[2,3,4], Yutaka Majima[5,*] and Ali Javey[1,2,6,*]

[1]Electrical Engineering and Computer Sciences, University of California, Berkeley, Berkeley, CA 94720, USA
[2]Materials Sciences Division, Lawrence Berkeley National Laboratory, Berkeley, CA 94720, USA
[3]The Molecular Foundry, Lawrence Berkeley National Laboratory, Berkeley, CA 94720, USA
[4]Department of Materials Science and Engineering, University of California, Berkeley, Berkeley, CA 94720, USA
[5]Materials and Structures Laboratory, Institute of Integrated Research, Institute of Science Tokyo, Yokohama, Kanagawa 226-8501, Japan
[6]Kavli Energy NanoScience Institute at the University of California, Berkeley, Berkeley, California 94720, United States

*Address correspondence to: majima.y.aa@m.titech.ac.jp, ajavey@berkeley.edu

**Tellurium (Te) is an intriguing one-dimensional (1D) semiconductor that has recently attracted considerable interest as a *p*-type channel material. However, scalable synthesis methods have lacked control over the orientation and patterning of the Te atomic chains, thus limiting its practical use. Guided by theory, we overcome this challenge using nanowire-shaped templates to achieve oriented, single-crystal growth of Te on amorphous substrates. Strong alignment of Te atomic chains is achieved as template widths are reduced to sub-20 nm. This high structural order, confirmed by 4D scanning transmission electron microscopy, enables the observation of pristine quantum transport phenomena for deterministically patterned Te. Field-effect transistors exhibit well-defined conductance plateaus at 77 K due to population of individual 1D subbands. Furthermore, Coulomb blockade emerges at 1.7 K, with the Te channel acting as a gate-tunable quantum dot. This synthesis approach provides a scalable pathway for integration of Te-based quantum materials for future electronic and quantum technologies.**

## 1 | Introduction

Tellurium (Te) is an anisotropic semiconductor that has recently attracted considerable interest as a *p*-type channel material due to its potential for large-area synthesis and excellent transport properties [1–7]. Te exhibits a 1D crystal structure composed of helical atomic chains bound by van der Waals interactions. Among conventional fabrication methods, the low-temperature evaporation of Te offers a promising route to wafer-scale uniformity, while being compatible with diverse integration platforms [1,8–10]. Te thin films produced by this approach undergo a subsequent amorphous-to-crystalline phase transition after deposition, resulting in a crystalline morphology with randomly oriented domains [11]. This material disorder is highly undesirable for device applications given its inherent variability and further limits the use of Te nanostructures as a scalable platform for fabrication of quantum materials with well-defined band structure [12–18].

Prior studies have examined substrate-induced alignment of Te on single-crystal MgO [19], mica [20], GaS and GaSe [21], m-plane sapphire using molecular engineering [22], as well as on three-fold symmetric metal dichalcogenides [23]. However, these approaches can be limited by unfavorable out-of-plane orientations or incomplete surface coverage. Furthermore, they generally rely on epitaxial substrates and high thermal budgets, which further limits the integration with existing Si technology. The preferential growth of Te nanowires and nanotubes along the helical chain axis is observed by high-temperature physical vapor deposition (PVD) with random coverage, poor thickness control and often necessitates transferring the structures from the growth substrate [24–33]. Liquid-phase self-assembly, on the other hand, provides excellent room-temperature alignment; however, it typically yields parallel-array networks where inter-wire junctions and residual ligands can influence transport properties. Consequently, a versatile, low-temperature method for achieving deterministic patterning and alignment of Te atomic chains remains to be established.

Here, we overcome this challenge by introducing a transfer-free patterning strategy that orients Te atomic chains into 1D quantum wires on an amorphous substrate via nanoscale templating. Drawing inspiration from the free energy minimization principles underlying block copolymer self-assembly [34,35], we use nanowire-shaped templates to facilitate edge nucleation and achieve precise alignment of single crystalline Te atomic chains along the template, as confirmed by four-dimensional scanning transmission electron microscopy (4D-STEM). Notably, strong alignment is observed along the template walls as feature widths are reduced to sub-20 nm. This method yields pristine, well-controlled quantum systems in Te. Specifically, population of individual 1D subbands is observed that manifest as clear conductance plateaus as a function of gate voltage at 77 K. Upon further cooling to 1.7 K, pronounced periodic oscillations are observed in the current-voltage characteristics due to Coulomb blockade. Our systematic investigation reveals how geometric confinement, specifically for wire widths below 20 nm, governs both the degree of atomic chain alignment and the energy-level discretization in the Te band structure. This approach provides a scalable pathway to unlock the intrinsic quantum properties of Te, paving the way for integration of Te into future quantum and advanced electronic technologies.

## 2 | Results and Discussion

### 2.1 | Atomic Chain Alignment Strategy

Physical vapor deposition of Te thin films onto cooled substrates yields a metastable amorphous phase, as limited adatom mobility suppresses crystallization. The subsequent crystallization of amorphous Te is a thermally activated process that proceeds with relatively low activation energy, allowing Te to crystallize near ambient conditions [11]. As shown in the schematic in Figure 1a, the atomic chain alignment begins with the thermal evaporation of Te onto a cooled substrate patterned with predefined nanoscale templates. At the low deposition temperature used in this study, the deposited film is amorphous, as the formation of a crystalline ground state is effectively blocked by kinetics. As the substrate temperature increases, nucleation of the crystalline phase becomes kinetically accessible. Note that this process differs from the driven nucleation and growth observed during typical gas phase deposition methods (e.g. molecular beam epitaxy, chemical vapor deposition) in that the crystalline phase is formed through a post deposition annealing, rather than during the deposition itself. In the presence of the template, nucleation of the crystalline phase from within the amorphous film can occur either heterogeneously within the wafer flat surface or at the template edges. We propose that heterogeneous nucleation at the template-substrate junction is favored, consistent with classical nucleation theory [36–38]. Once the nuclei exceed the critical size, crystal growth proceeds more rapidly along the *c*-axis rather than in other directions [11,23]. As a result, Te atomic chains that are in contact with the template walls become aligned along the template direction, dictated by the junction geometry. In contrast, regions farther from the walls experience less influence from the template, leading to a competition between wall-induced alignment and random orientation. Thus, the degree of atomic chain alignment within the template is determined by the balance between edge-nucleated and directed growth and nucleation events in the film interior, suggesting that the template width exerts a strong influence on the alignment of Te atomic chains. Furthermore, a low growth temperature is essential to kinetically slow the amorphous-to-crystalline phase transition. This suppressed kinetic environment minimizes random nucleation events and improves overall film uniformity, ensuring that the template-guided chain alignment dominates.

## 2.2 | Evaluation of Structural Anisotropy

The template material plays a key role in dictating the surface energy landscape and free energy minimization pathway, which in turn govern preferential heterogeneous nucleation and crystal growth. In this work, we employed poly(methyl methacrylate) (PMMA) as the template material that can be deterministically patterned via electron beam lithography to align the Te atomic chains into arrays of quantum wires. Structural characterization was performed using transmission electron microscopy (TEM) with the electron beam perpendicular to the substrate, after crystallization and removal of the polymer template (Figure 1b, see Experimental Section). TEM imaging and selected-area electron diffraction (SAED) acquired from individual quantum wires indicate single-crystalline Te. While annular virtual dark-field images exhibit spatial bright-dark contrast, this intensity variation originates from local diffraction conditions, such as minor lattice strain and dynamical electron scattering, and does not necessarily indicate distinct crystallographic domains. Consequently, crystallographic orientation was determined using mean diffraction patterns sampled at various positions along the Te wires. The SAED patterns can be indexed such that the [0001] crystallographic direction, corresponding to the $c$-axis of the Te helical chains, is aligned with the template-defined nanowire axis (Figure 1b and Figure S1). The diffraction spots were analyzed to extract lattice parameter ($c$ = 5.932 Å) consistent with trigonal Te. High-resolution TEM (HRTEM) imaging of a narrow Te quantum wire (width, $w$ = 16 nm and height, $h$ = 6 nm) shows the screw-like atomic chains of Te, as highlighted by the overlaid atomic structure (Figure S2). The fast Fourier transform (FFT) further confirms alignment of the [0001] crystallographic direction with the template direction (Figure S2a, inset).

The structural anisotropy of aligned Te quantum wire was investigated using angle-resolved polarized Raman spectroscopy. To increase the measured signal, Te was patterned into parallel arrays of quantum wires with a pitch ranging from 100 to 200 nm. (Figure 1c, d). Three expected vibrational modes in the Raman spectrum of Te are observed at 93.8 $cm^{-1}$ ($E_1$ mode), 123.6 $cm^{-1}$ ($A_1$ mode) and 141.2 $cm^{-1}$ ($E_2$ mode) (Figure S3a). The $E_1$ ($E_2$) mode originates from asymmetric stretching along the [0001] direction, while the $A_1$ mode corresponds to the symmetric expansion of Te atoms in the basal plane [7]. The small blue-shift of these vibrational frequencies is attributed to reduced van der Waals coupling in the Te thin film [3]. The $A_1$ mode exhibits greater polarization sensitivity relative to the $E_1$ mode in aligned quantum wires, reaching an anisotropy ratio of ~ 6.6 as the polarization angle varies from 0° to 90°, as shown in Figure 1e and Figure S3a. Notably, the $A_1$ mode intensity reaches a minimum at 0° and 180°, when the incident light polarization is aligned perpendicular to the *c*-axis of Te. This behavior is consistent with theoretical predictions based on the Raman polarizability tensors and the deformation potential associated with electron-phonon coupling [39].

As a control, angle-resolved polarized Raman measurements were also performed on an unpatterned Te thin film of similar thickness (hereafter referred to as "bulk") (Figure S3c, d). In contrast to the aligned quantum wires, no significant polarization-dependent modulation of Raman intensity is observed in the bulk thin film. This result is consistent with previous reports [1], resulting in an averaged Raman response from multiple randomly oriented chains from adjacent grains and thereby suppressing anisotropic signatures.

Quantitative assessment of atomic chain alignment requires detailed analysis of the electron diffraction patterns. To achieve this, we used 4D-STEM to acquire spatially resolved diffraction data along the length and width of the quantum wire. Diffraction patterns extracted at discrete probe positions along the wire reveals variation in the zone axes, with certain orientations exhibiting stronger diffraction intensities (Figure S4a, b). These intensity differences arise from the orientation-dependent scattering efficiency of the electron beam. When the incident beam aligns with a high-symmetry zone axis, where atomic planes are more regularly spaced and periodically arranged along the projection direction, the resulting diffraction condition more strongly satisfies the Bragg condition across multiple planes, leading to constructive interference and enhanced diffraction spot intensity. Conversely, off-axis orientations reduce the coherence of scattered waves, diminishing the diffraction contrast. This behavior of experimental diffraction patterns is consistent with simulated diffraction patterns for trigonal Te, which show minimal variation in the *c*-axis orientation, corresponding to the [0001] direction, across different regions of interest along the wire (Figure S4c, e). The gradual variation of diffraction patterns and zone axis orientation along the wire, despite the preserved [0001] atomic chain direction, suggests a continuous lattice rotation consistent with long-range helicity or twist. A more detailed crystallographic analysis will be necessary to quantitatively characterize this structural feature in future work. The spatially resolved orientation angles of the Te atomic chains, extracted from each pixel within the measurement window, along with the mean diffraction pattern of the entire dataset, confirm that the Te atomic chains maintain a single-orientation crystalline structure throughout the template (Figure S5).

Identifying the crossover template feature size at which alignment of Te atomic chains begins to deteriorate is essential for achieving precise structural control. We define the atomic chain orientation angle ($\theta$) as the angular deviation between the template direction and the [0001] crystallographic axis of the Te helical chains (Figure 2a). This angle is determined for each quantum wire using diffraction data and calibrated TEM images. Figure 2b presents discrete measurements of $\theta$ as a function of template width ($w$), revealing that Te atomic chains become increasingly well-aligned as the template width decreases. This trend underscores the critical role of nanoscale confinement in governing chain alignment within patterned templates. To further evaluate alignment uniformity, we plotted the mean orientation angle ($\bar{\theta}$) over micrometer-scale lengths for various pattern widths (Figure S6). Notably, because no domain boundaries were detected within these measured regions, we found no evidence that the template width limits the maximum continuous length of the single-crystalline domains. Instead, the primary role of the template width is strictly governing the degree of crystallographic alignment of these micron-scale Te crystals. For example, at template widths below 20 nm, $\bar{\theta}$ remains below 5°, supporting our hypothesis that edge nucleation dominates in narrow templates. As the template width increases, $\bar{\theta}$ can become larger, although the standard deviation within each template remains small. This confirms that the broad scatter of orientation angles observed at larger template widths in Figure 2b represents statistical variations between distinct wires, rather than the formation of polycrystalline domains within a single wire. Because wider templates increase the probability of random nucleation on the flat wafer surface, these distinct, single-crystalline wires can nucleate and grow at any arbitrary angle. Consequently, the broad inter-wire distribution observed at these large dimensions is a direct manifestation of the statistical competition between highly aligned edge nucleation and randomly oriented flat-surface nucleation.

To preserve the intrinsic atomic chain orientation and avoid structural damage often induced by substrate-to-membrane transfer processes, all TEM samples were fabricated directly on SiN membranes via electron-beam lithography (see Experimental Section). Furthermore, while edge roughness in lithographic templates can adversely affect the growth orientation of emerging nuclei, our template remains smooth following electron-beam exposure. As a result, the edge roughness observed in the final Te wires (Figure S6b) is primarily induced by the mechanical lift-off procedure. Because this roughening occurs after crystallization, it has a negligible impact on the initial nucleation and alignment phases.

A classical nucleation model provides framework for interpreting these experimental observations. Assuming that the crystallization kinetics of Te on amorphous SiN are comparable to those on $SiO_2$ [11], the patterned regions are sufficiently small that, for a domain 1.5 μm in length, only a single nucleus is expected to form. Consequently, the final orientation of the aligned chains is determined by the orientation of this initial nucleus. The distribution of atomic chain orientations is therefore governed by the relative rates of two competing heterogeneous nucleation processes: nucleation at the template-substrate junction (edge nucleation) and nucleation on the flat wafer surface within the patterned template (flat-surface nucleation).

The orientation of the nuclei forming on the flat wafer surface is assumed to be uniformly distributed between 0 and 90°. Conversely, the orientation of nuclei forming at the template-substrate junction is assumed to follow a Gaussian distribution. The final probability distribution for the orientation of the atomic chains is given by,

$$P(\theta) = \left( \frac{1 - f_{\text{edge}}}{90} + f_{\text{edge}} \sqrt{\frac{2}{\pi\sigma^2}} e^{-\frac{\theta^2}{2\sigma^2}} \right) d\theta \tag{1}$$

where $f_{\text{edge}}$ is defined to be the fraction of nuclei that form at the template-substrate junction, and $\sigma$ is the parameter governing the width of the Gaussian. In deriving (1), we have assumed that $\sigma \ll 90°$. Physically, the parameter $\sigma$ is meant to reflect the roughness in the PMMA template. It is assumed that nuclei will form with their *c*-axis aligned along the local tangent of the template. Then the average angle, $\langle\theta\rangle$, depends upon the parameters of the distribution according to:

$$\langle\theta\rangle = \left(1 - f_{\text{edge}}\right) 45 + f_{\text{edge}} \sqrt{\frac{2}{\pi}} \sigma \tag{2}$$

with the root mean square deviation given by:

$$\langle\theta^2 - \langle\theta\rangle^2\rangle^{1/2} = \left[2700\left(1 - f_{\text{edge}}\right) + f_{\text{edge}}\sigma^2 - \left(45 + f_{\text{edge}}\left(-45 + \sqrt{\frac{2}{\pi}}\sigma\right)\right)^2\right]^{1/2}. \quad (3)$$

The fraction $f_{\text{edge}}$ can be expressed in terms of the length defined by the ratio of the edge nucleation rate to the flat-surface nucleation rate, $\lambda = \nu_{\text{edge}}/\nu_{\text{flat}}$,

$$f_{\text{edge}} = \frac{2\lambda}{w + 2\lambda}, \quad (4)$$

with $w$ the template width and the factors of 2 in (4) stemming from the fact that there are two edges available for nucleation. By combining equations (2) and (4), the average orientation angle can be modeled as a function of $\lambda$ and $w$, while equations (3) and (4) provide an estimate of the intrinsic spread in $\theta$. Figure 2c presents the experimentally measured average orientation angle, binned by template width in 5 nm intervals. Fitting the data to equations (2) and (4) yields $\sigma =$ 1.87° and $\lambda$ = 62 nm, with the uncertainty corresponding to a 95% confidence interval. If one defines the crossover template width to be the sample width at which one expects $f_{\text{edge}} = 1/2$, the crossover width at 278.15 K is ~124 nm. The fitted model, overlaid with the experimental data in Figure 2c, shows good agreement given the intrinsic width of the probability distribution.

### 2.3 | Quantum Wire Transistors from Aligned Chains

Te is a promising channel material for *p*-type field effect transistors (FETs), owing to its high hole mobility in the bulk limit and ambient stability. To harness the full potential of Te for nanoscale electronics, it is critical to evaluate the electrical properties of ultra-thin, aligned Te quantum wires. Figure 3a illustrates the schematic of a back-gated FET, in which aligned Te atomic chains bridge the source and drain electrodes. A global back-gate is used to modulate the carrier density and nickel is chosen as the contact metal due to favorable work function, ensuring low-resistance ohmic contact with *p*-type Te. The output characteristics ($I_d - V_d$) in Figure 3b demonstrate that the quantum wires can sustain substantial current, delivering an on-state current of approximately 190 μA/μm at room temperature. The transfer characteristic ($I_d - V_g$) for the same wire is shown in Figure 3c. Devices fabricated using the templated alignment method exhibit good switching behavior, with an average $I_{max}/I_{min}$ current ratio exceeding four orders of magnitude. Notably, the narrow quantum wires also achieve high field-effect mobilities, with a peak value reaching 60.9 $cm^2/Vs$ as shown in Figure 3d.

It should be noted that at the aggressively scaled dimensions of these Te quantum wires, carrier mobility is fundamentally limited by surface and edge roughness, which induce severe spatial fluctuations in the band edge potential, as well as scattering from interface defects. Therefore, to quantitatively benchmark device performance, we compared the maximum width-normalized on-state current of our Te quantum wires against previously reported Te nanowires and nanosheets of comparable nanoscale dimensions. As shown in Figure S7, the normalized on-state current of the highly oriented quantum wires improves dramatically compared to unpatterned evaporated thin films deposited under identical conditions, underscoring the electronic advantages of the template-aligned single-crystalline chains. Furthermore, large-area scalability of this deterministic patterning, deposition and crystallization strategy was demonstrated by fabricating arrays of Te quantum wire devices across a full 4-inch wafer. As shown in Figure S8, these devices exhibit consistent electrical transport properties at the wafer scale.

Although Te can oxidize at room temperature, the surface oxidation is minimized by using a thin ALD-deposited $HfO_2$ capping layer. Time-dependent Raman and electrical transport measurements demonstrate stability to ambient exposure over several weeks (Figure S9).

Additionally, the thermal stability of the quantum wires was evaluated by stepwise thermal annealing at 55°C and 85°C, after which the drain current showed negligible change (Figure S10). This highlights the robustness of the wires under technologically relevant operating temperatures.

While most investigations on low-dimensional Te have focused on thin films [40,41], where quantum confinement is achieved solely by thickness reduction, probing the 1D band structure in Te requires additional lateral scaling to confine the helical chains into aligned quantum wires and fully exploit the material's structural anisotropy. In particular, high temperature PVD typically yields Te nanowires and nanosheets [42,43] with dimensions significantly larger than the exciton Bohr radius of Te, where quantum confinement effects remain negligible. In contrast, our templated alignment deterministically places high-quality Te quantum wires with lateral dimensions below 20 nm and thicknesses under 10 nm, which are small enough to induce strong quantum confinement and quantized subband formation. This enables 1D transport measurements at cryogenic temperatures.

A 1D Te quantum wire can be modeled as a particle-in-a-box in the transverse directions, where carriers are tightly confined by an effective potential well in two dimensions, giving rise to discrete subband energies, while remaining free to move along the wire axis. At sufficiently low temperatures, where phonon-induced scattering is strongly suppressed, the controlled population of these discrete subbands become observable, manifesting as conductance plateaus in electrical measurements. Note that other scattering sources such as structural imperfections or surface states and finite thermal broadening can obscure these 1D transport signatures, making quantized plateaus difficult to resolve. Accordingly, in our two-terminal configuration the plateau heights are also limited by the effective series contact resistance and the transmission coefficient, so the observed plateaus reach only a small fraction of the quantum conductance $G_0$ at these temperatures (Figure S11). Figure 4a shows the transfer characteristics of a representative aligned Te quantum wire at various temperatures. Once the device is cooled below approximately 100K, clear plateaus appear in the $I_d$-$V_g$ trace.

The number of resolvable conductance plateaus in the transfer characteristics depends on both electrostatic gate-coupling efficiency and the geometric confinement set by the wire's cross-section. Figure 4b compares $I_d$-$V_g$ curves for three aligned quantum wires of identical thickness (6 nm) but varying widths. For the 13 nm and 18 nm wires, two subband-derived plateaus appear within the gate-voltage sweep range, since the energy separation $\Delta E_{21}$ between the first and second subbands is large enough to resolve. As the channel width increases beyond ~20 nm, $\Delta E_{21}$ shrinks, owing to weaker confinement, so that the additional plateaus from higher subbands emerge at closely spaced voltages, making them difficult to distinguish. In earlier work on cylindrical nanowires, symmetric confinement produced two-fold degeneracy in certain subband energies [13,17]. However, for wires with a rectangular cross-section, such as those in the present study, the subbands are non-degenerate, with each exhibiting a distinct energy level. This characteristic is confirmed by solving Schrödinger equation for a 2D rectangular potential well (Figure 4c). This behavior persists across the range of widths studied, allowing us to clearly resolve each subband in the aligned quantum wires. Figure 4d plots the gate voltage difference $\Delta V_{21}$ between subband (2,1) and (1,1), extracted from peaks in the transconductance $g_m$, for multiple wires ranging from 12 nm to 22 nm in width (Figure S12).

The observed subband energy scaling is further supported by the analytic solution of a 2D infinite-barrier potential well. Assuming that the valence-band maximum of Te is separated from the surrounding dielectric/ambient medium by an effectively infinite barrier (Figure S13), the Schrödinger equation in the wire cross-section with a height $h$ and width $w$ becomes,

$$-\frac{\hbar^2}{2m_\mathrm{h}^*}\nabla^2\psi(x,y)+V(x,y)\psi(x,y)=E_{\mathrm{m,n}}\psi(x,y) \tag{5}$$

$$E_{\mathrm{m,n}}=\frac{\hbar^2\pi^2}{2m_\mathrm{h}^*}\left(\frac{m^2}{w^2}+\frac{n^2}{h^2}\right) \tag{6}$$

where $m_\mathrm{h}^*$ is the effective hole mass in Te, $\nabla^2$ is the 2D Laplacian operator, $\psi(x,y)$ is the transverse wave function, $V(x,y)$ is the potential in the wire cross-section and $E_{\mathrm{m,n}}$ is the eigen energy of the (m,n) subband. In Figure 4d, the measured voltage spacing $\Delta V_{21}$ exhibits an inverse-square dependence on the wire width $w$, consistent with the analytical prediction $\Delta E_{21} = E_{2,1} - E_{1,1} \propto 1/w^2$. Different dielectrics (e.g., $SiO_2$, SiN, and high-κ oxides) can have disparate surface chemistries, which could in principle influence Te crystallization kinetics and chain alignment. To test whether the deterministic nature of the growth is substrate-independent, we extended the electrical measurements beyond the original dielectric system. In addition to $SiO_2$, we fabricated Te quantum wires on $HfO_2$ and $ZrO_x$, where the transfer characteristics exhibit subband features consistent with 1D quantum transport (Figure S14). These results show that the template-directed alignment and electronic structure are not limited to a single amorphous dielectric platform. Furthermore, to decouple the effect of the polymer template from Te crystallization, we repeated the patterning and alignment using ZEP 520A as an alternative template. The resulting devices show subband plateaus consistent with those obtained using PMMA (Figure S15), indicating that the template choice does not qualitatively alter the alignment process.

Upon further cooling of the aligned quantum wire devices to 1.7 K, the subband-derived plateaus evolve into pronounced periodic oscillations in gate voltage, with an average spacing of $\Delta V_\mathrm{g}$ = 130 mV (Figure 4e). This behavior is characteristic of Coulomb blockade, a phenomenon well-documented in carbon nanotubes and semiconductor quantum wires [44–47]. Coulomb blockade occurs when a quantum-confined "island" is weakly coupled to its contacts, either because of structural disorder or significant tunnel barriers, so that adding a single charge requires a charging energy $\Delta E_N = e^2/C_\Sigma$ that exceeds the thermal energy $k_B T$. In that regime, charge transport is blocked except when the Fermi level in one of the contacts aligns with an empty single-particle state in the island. Figure 4f illustrates this schematically, where all single-particle levels $E_N$ in the island that lie above the source Fermi level $E_{F,s}$ are occupied by holes. Transport becomes possible only if the drain Fermi level $E_{F,d}$ aligns with the next unoccupied level $E_{N+1}$, i.e $E_{F,d} < E_{N+1} < E_{F,s}$. Each occurrence of the alignment produces a conductance peak ($G = dI/dV$) in the Coulomb oscillations. Between the peaks, the conductance is suppressed because the next available level exceeds the contact Fermi energies. Equivalently, at fixed drain bias one can sweep the back-gate voltage to modulate the island's potential and observe Coulomb oscillations. The absence of any resolvable excited-state features in Figure 4e indicates that the single-particle levels in the Te wire are roughly equidistant, with no detectable excited-level splitting within the charging regime.

To pinpoint the origin of these Coulomb oscillations, we performed bias-spectroscopy measurements on an aligned quantum wire. Figure 4g shows the differential conductance map as a function of gate voltage and source-drain bias. Periodic Coulomb diamonds are observed that signify single-hole tunneling through a quantum-confined island, from which we extract a gate capacitance of $C_g = e/\Delta V_g = 1.23$ aF. Analysis of the diamond's full extent yields a self-capacitance of $C_\Sigma = 54.7$ aF, corresponding to a charging energy $\Delta E_N = 2.92$ meV. This is consistent with the maximum bias $V_{ds,max}$, delineated by the diamond boundaries (Figure S16). To compare this to an analytic estimate, we model the wire as a cylindrical conductor above a dielectric plane with,

$$C_\Sigma = \frac{2\pi\epsilon_0\epsilon_r L}{\ln\left(\frac{2t_{ox}}{R}\right)} \tag{7}$$

Here, $\epsilon_r$ is the permittivity of the surrounding dielectric, $L$ is the length of the isolated island, $t_{ox}$ is the oxide thickness and $R$ is the effective radius of a circle having the same area as the wire's cross-section [48]. For $L$ = 500 nm with an effective radius of 7 nm, Equation (7) yields $C_\Sigma$ = 40.8 aF, in reasonable agreement with the capacitance obtained from the Coulomb diamond. This implies, the entire 500 nm length of the wire contributes to the island's self-capacitance, with tunnel barriers formed at the metal-Te interfaces.

## 3 | Conclusion

We have demonstrated a simple, template-driven method to align Te atomic chains without the use of epitaxy. This approach opens unprecedented opportunities for exploring self-aligned Te nanostructures, ranging from quantum wires to quantum dots. In the future, strain may be engineered by designing the shape of templates into nano-rings, providing valuable insights into strain-tunable quantum phenomena. Furthermore, controlling the crystallization of Te atomic chains in zero-dimensional dots could enable the design of single-electron transistors, where both the leads and the dot are composed of Te and fabricated in one step, thereby mitigating material disparity and enhancing device integration. Finally, given the low thermal budget of the fabrication process and the relatively high hole mobility of Te, monolithic 3D integration of single-crystalline Te nanowires could be explored with Si CMOS for future AI hardware.

## 4 | Experimental Section

**4.1 | Templated growth of aligned Te atomic chains.** Nanoscale templates were defined on commercially available 3 mm × 3 mm SiN chips (10 nm membrane thickness on a 200 μm silicon frame). Poly(methyl methacrylate) resist was spin-coated onto the SiN membrane and baked at 180°C for 5 minutes. Electron-beam lithography (Raith EBPG5200) was used to pattern the resist

into templates, followed by development in a 1:3 mixture of methyl isobutyl ketone and isopropanol. High purity Te pellets (99.999%, Sigma-Aldrich) were thermally evaporated in an Edwards E306A Coating System equipped with a custom cryogenic stage. The patterned substrates were evacuated to a base pressure of $1\times10^{-6}$ Torr, then cooled to 193 K by flowing cold nitrogen gas. Te was deposited at a rate of 4 Å/s and the thickness of the film was monitored using a quartz crystal microbalance. Film thickness was subsequently confirmed by atomic force microscopy (Bruker Icon Dimension with ScanAsyst). After deposition, the substrate was allowed to warm near room temperature under nitrogen flow, and the Te film was crystallized by holding the sample at 278 K. Lift-off was performed by immersing the samples in acetone with gentle agitation achieved via the uniform flow generated by a rotating magnetic stir bar. No additional agitation was applied.

**4.2 | TEM and Raman characterization.** TEM measurements were conducted at the National Center for Electron Microscopy (NCEM), Lawrence Berkeley National Laboratory, using a FEI TitanX 60-300 operating at 300 kV. For 4D-STEM experiments, the microscope was operated in STEM microprobe mode at 300 kV, utilizing a 40 µm C2 aperture, spot size 10, a camera length of 195 mm, and an indicated convergence semi-angle of 0.48 mrad. Nanodiffraction datasets were acquired using a Gatan Orius 830 camera with a dwell time of 100 ms per probe position. Data processing and image analysis were carried out using the open-source software package py4DSTEM [49], supplemented with custom Python scripts. Parallel arrays of aligned Te quantum wires were fabricated on 50 nm $SiO_2/p^+$ Si substrates, using the nanoscale-template method detailed above, to enhance light-matter coupling along the template axis. Raman spectra were acquired on a Horiba LabRAM HR Evolution confocal microscope. A 473 nm solid-state laser

was used as the polarized light source. A zero-order half-wave plate in the excitation path allowed continuous rotation of the incident polarization relative to the template axis. A long integration time was employed to maximize the signal-to-noise ratio of the collected spectra.

**4.3 | Device fabrication and measurement.** Individual Te quantum wires were defined on a $SiO_2/p^+$ Si substrate with 50 nm $SiO_2$ as the gate dielectric and the highly doped silicon as the global back-gate. Source and drain electrodes were patterned by electron-beam lithography, followed by thermal evaporation of 20 nm Ni and lift-off in acetone. A thin layer (~ 3nm) of ALD deposited $HfO_2$ is used as a capping layer to minimize surface oxidation. Electrical measurements at both room temperature and 77 K were carried out under vacuum using a Lakeshore cryogenic probe station equipped with a Keysight 4155C semiconductor parameter analyzer, while cryogenic transport down to 1.7 K was measured in a Quantum Design PPMS DynaCool. Field effect mobility ($\mu_{FE}$) was extracted from the linear regime using,

$$\mu_{\mathrm{FE}} = g_{\mathrm{m}} \frac{L^2}{C_{\mathrm{ox}}} \frac{1}{V_{\mathrm{ds}}}$$

where, $g_m = dI_d / dV_g$ is the transconductance, $L$ is the channel length and $C_{ox}$ is the total gate oxide capacitance. $C_{ox}$ was obtained from 2D AC mixed-mode simulation in Sentaurus TCAD. In the quasi-1D regime, where the width of the quantum wire is narrower than the gate oxide thickness, the standard parallel-plate capacitance approximation breaks down. Electrostatic fringing fields extending from the sidewalls of the nanowire to the bottom gate begin to dominate the overall capacitive coupling. The TCAD simulation accurately captures these pronounced edge effects, providing a correct estimation of the effective capacitance (Table S1).

**Acknowledgments**

This work was funded by the U.S. Department of Energy, Office of Science, Office of Basic Energy Sciences, Materials Sciences and Engineering Division under Contract No. DE-AC02-05-CH11231 (EMAT program KC1201). Work at the Molecular Foundry was supported by the Office of Science, Office of Basic Energy Sciences, of the U.S. Department of Energy under Contract No. DE-AC02-05CH11231. Y.M. acknowledges funding from JST CREST (JPMJCR22B4) and Data Creation and Utilization Type Material Research and Development, MEXT (JPMXP1122683430), which funded parts of the electron beam lithography patterning. N.H. acknowledges support from JST PRESTO (No. JPMJPR23H7), Japan, for the material supplies.

**Author Contributions**

IKMRR, YM and AJ conceived the idea for the project and designed the experiments. IKMRR, SD, SV, GO, AN, IK and HMK fabricated devices. IKMRR, NH, AK and GO performed electrical and optical measurements. MJ and KCB performed TEM characterization. IKMRR, TK, NH, KCB, JWA, MS, YM and AJ analyzed the data. DCC performed theoretical modeling. IKMRR, MJ, NH and AJ wrote the manuscript. All authors discussed the results and commented on the manuscript.

**Conflict of Interest**

The authors declare no conflicts of interest.

**Data Availability Statement**

The data that support the findings of this study are available from the corresponding author upon reasonable request.

**Corresponding Author**

*E-mail: majima.y.aa@m.titech.ac.jp, ajavey@berkeley.edu

**Figures**

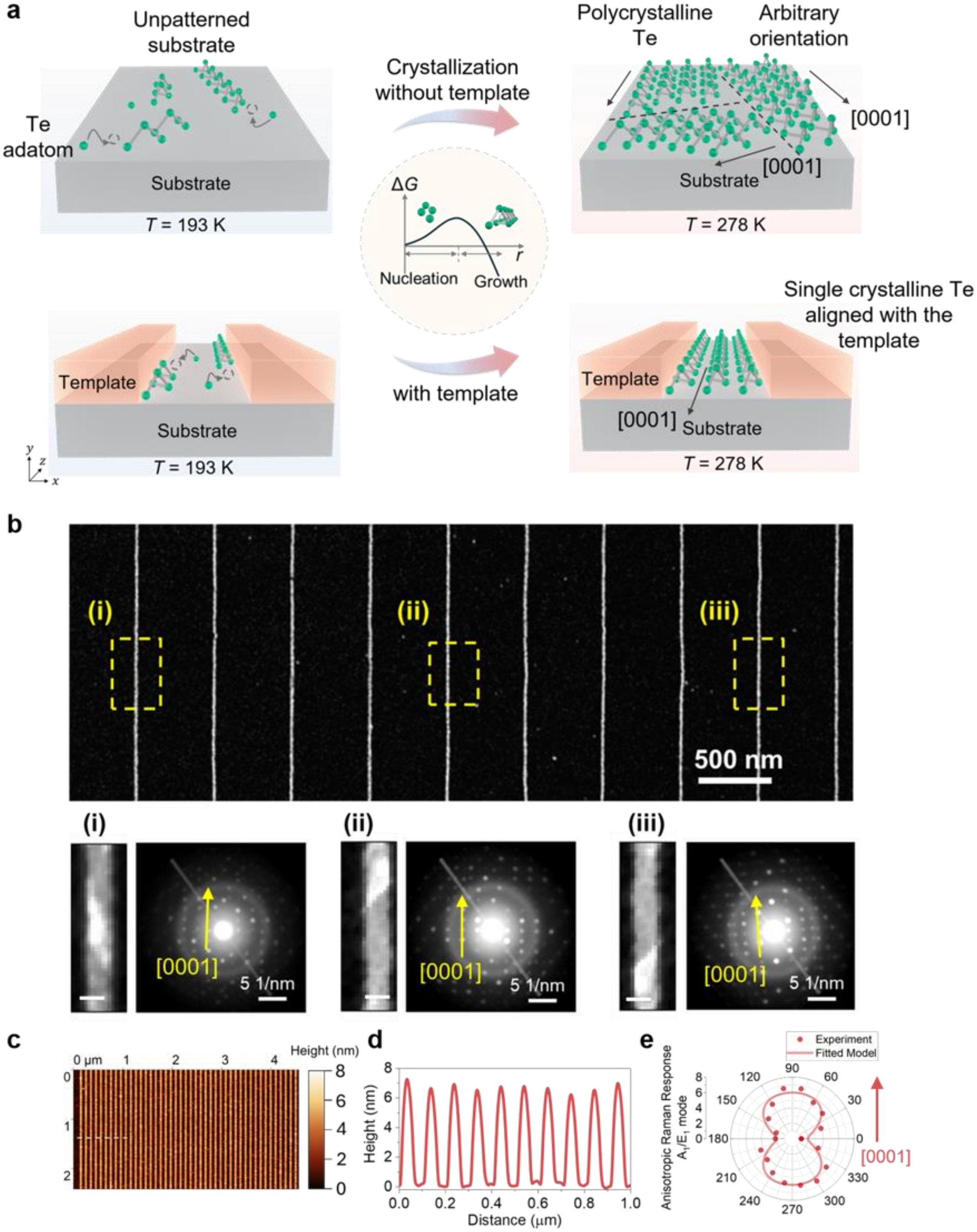

**Figure 1 |** Alignment of Te atomic chains. (a) Schematic illustration of the alignment strategy using a nanowire-shaped template. Te is thermally evaporated onto a substrate cooled to 193 K, where limited adatom mobility results in a metastable amorphous film. Upon subsequent warming

to near-ambient temperature (278 K), Te undergoes an amorphous-to-crystalline phase transition. (Top) On an unpatterned substrate, homogeneous nucleation leads to randomly oriented atomic chains and the formation of multiple crystalline domains. (Bottom) In contrast, with a nanowire-shaped template present, heterogeneous nucleation occurs preferentially at the junction between the template and substrate. Once nuclei exceed the critical size, crystal growth proceeds more rapidly along the $c$-axis than in other directions, thus governing the atomic chain alignment within the template. (b) (Top) TEM image of deterministically patterned Te quantum wires on a SiN membrane. The shaded boxes indicate the regions of interest where diffraction patterns were measured. (Bottom) Virtual dark-field image of a Te quantum wire reconstructed from a 4D-STEM scan, alongside the corresponding diffraction patterns for the three dashed regions (i–iii). The arrow indicates the [0001] crystallographic direction (the $c$-axis of the helical Te chains). Scale bar: 20 nm. (c) Atomic force microscopy (AFM) micrograph showing deterministically aligned Te atomic chains patterned by the PMMA template, following crystallization and lift-off. (d) AFM height profile along the dashed white line indicated in (c), showing an average Te thickness of 6 nm. (e) Polar plot of the Raman intensity ratio $I_{A1}/I_{E1}$ for an array of quantum wires.

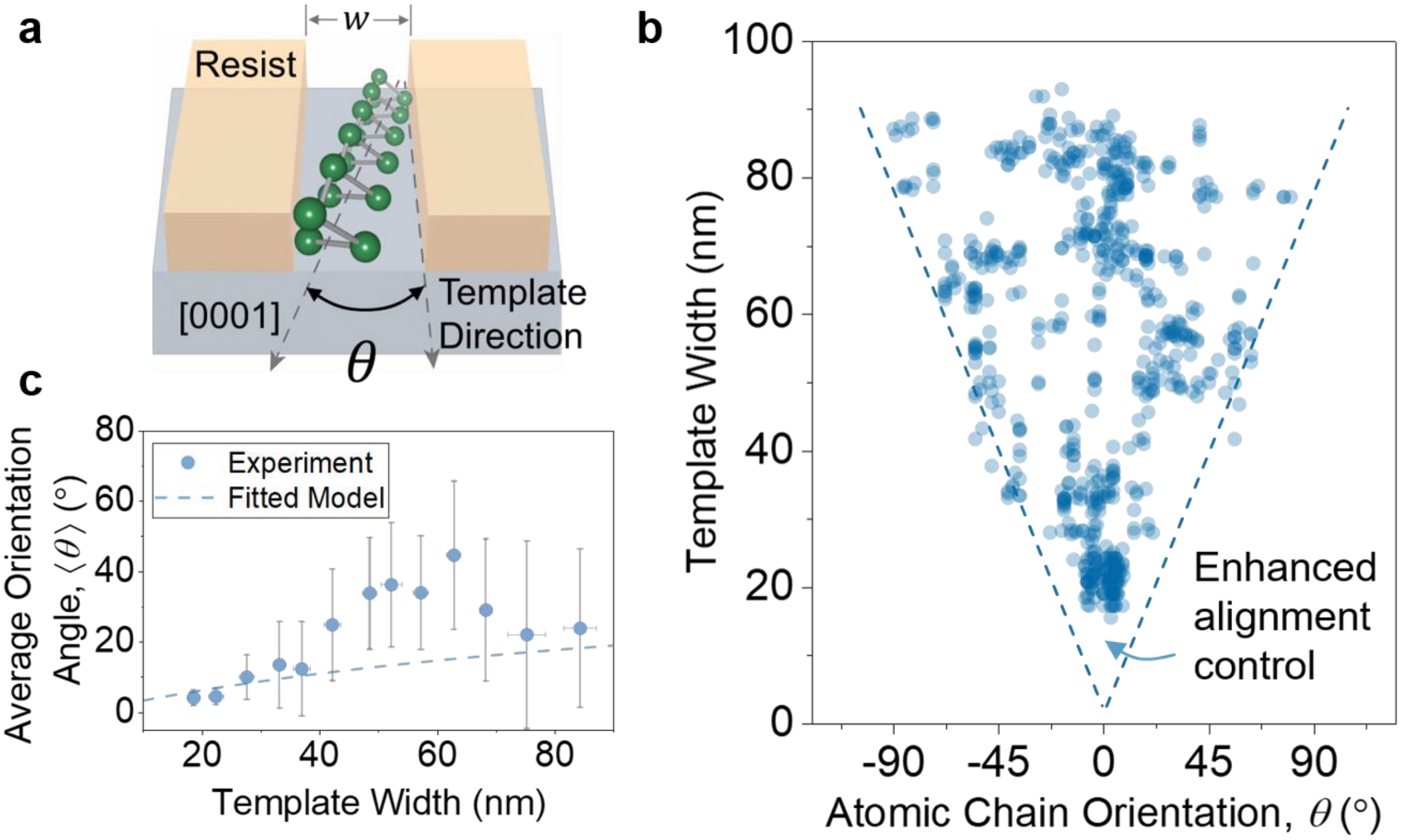

**Figure 2 |** Alignment control as a function of template width. (a) Schematic illustration of the atomic chain orientation angle ($\theta$) which is defined as the angle between the template direction and the [0001] crystallographic axis of the Te helical chains. (b) Atomic chain orientation angle ($\theta$) as a function of template width ($w$), demonstrating enhanced alignment control in narrow templates. The dashed lines serve as a visual guide to indicate the spread of the data. (c) Mean orientation angle $\langle\theta\rangle$ versus template width, comparing experimental data with the fitted model from equation (2).

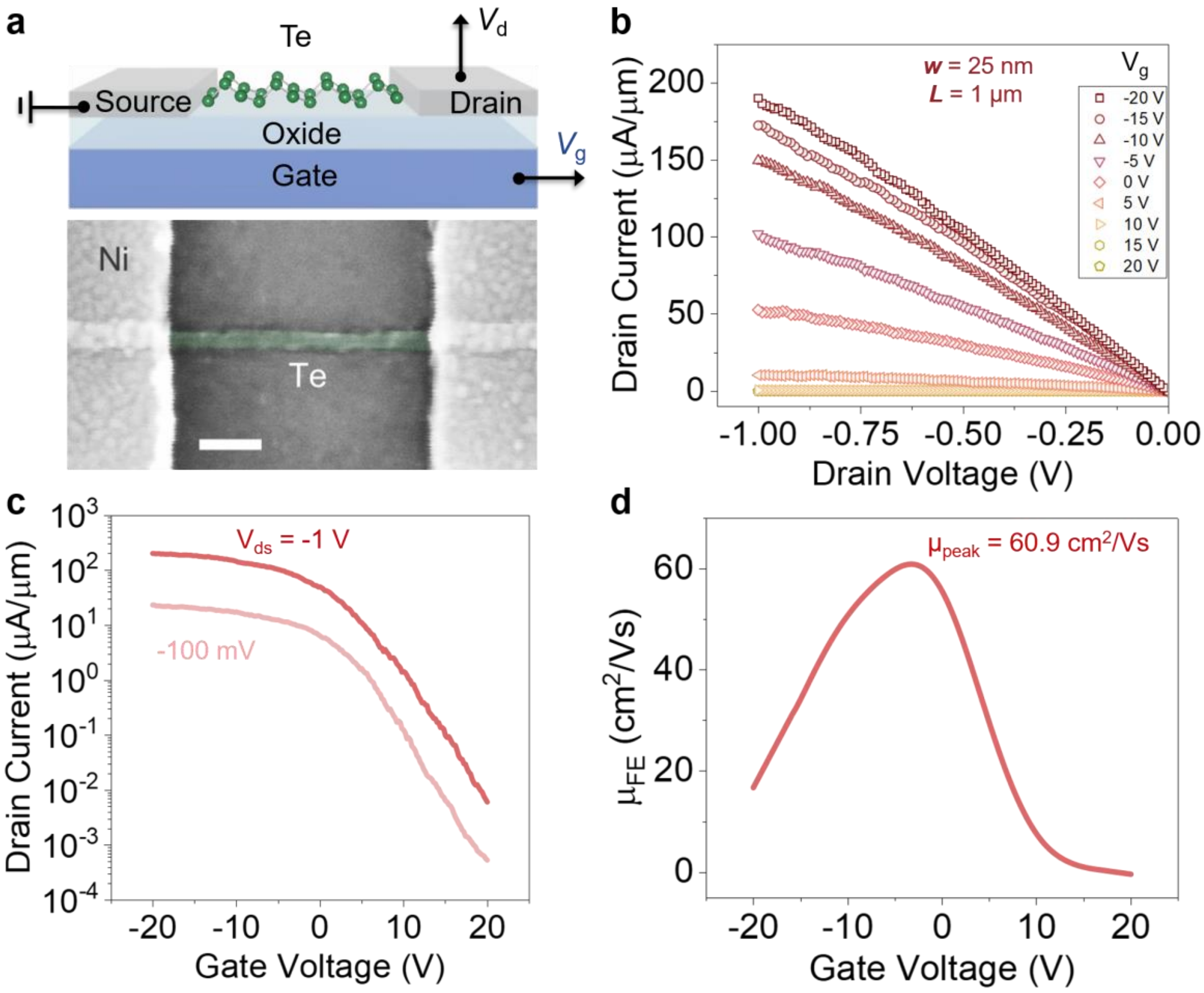

**Figure 3 |** Room-temperature electrical performance of aligned Te quantum wires. (a) (Top) Schematic of a back-gated FET with the atomic chains bridging the source and drain electrodes. (Bottom) False color SEM image of a representative device. Scale bar: 50 nm. (b) Output characteristics ($I_d - V_d$) of a quantum wire FET with a channel width $w$ = 25 nm and length $L$ = 1 μm on $SiO_2$ gate dielectric, exhibiting high on-state current. (c) Transfer characteristic ($I_d - V_g$) of the same device at different drain biases. (d) Extracted field-effect mobility as a function of gate voltage at $V_{ds}$ = -1 V. Thickness of Te is 6 nm.

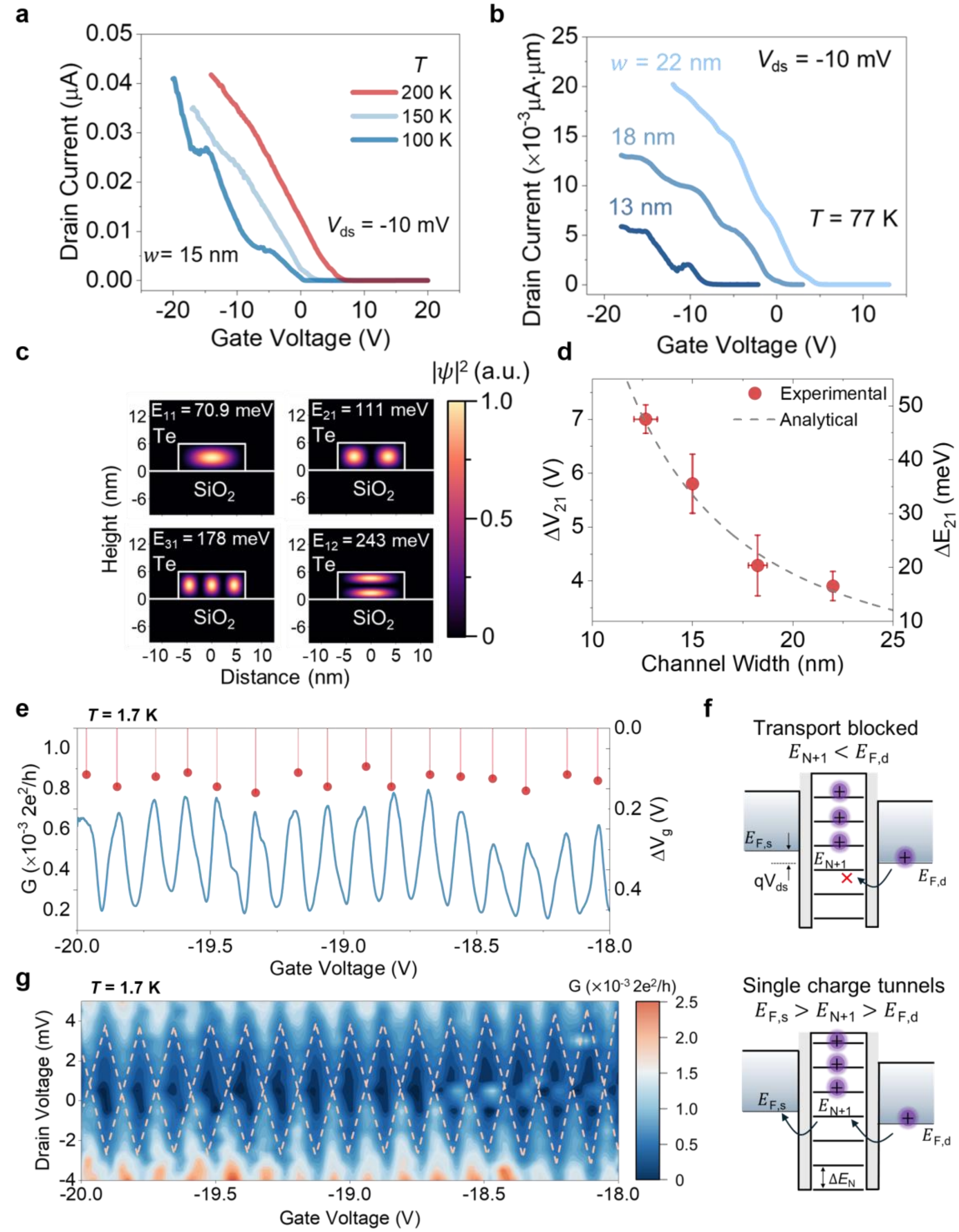

**Figure 4 |** Transport characteristics of Te quantum wires on $SiO_2$ dielectric. (a) Transfer characteristics of a representative Te quantum wire measured at various temperatures. The $I_d$-$V_g$ curves for $T = 100$ K has been intentionally offset by +3 V for improved clarity. (b) Transfer curves

for Te quantum wires with varying channel widths (constant thickness of 6 nm and length $L$ = 200 nm), highlighting quantum confinement effects. The $I_d$-$V_g$ curves for $w$ = 22 nm has been intentionally offset by +8 V for improved clarity. (c) Calculated hole probability density $|\psi|^2$ across the wire cross-section, obtained from numerical solution of the two-dimensional Schrödinger equation with infinite barrier confinement. (d) Extracted energy difference between the first and second subband versus wire width, showing inverse scaling behavior. (e) Periodic Coulomb oscillations observed at 1.7 K in an aligned Te quantum wire ($L$ = 500 nm), with equidistant conductance peaks ($\Delta V_g$) indicating the absence of resolvable excited states. (f) Energy-level schematic of Coulomb blockade in a Te quantum wire, showing tunneling barriers at the metal–semiconductor interfaces. (g) Differential conductance map as a function of gate voltage and source-drain bias, displaying periodic Coulomb diamonds characteristic of single-charge tunneling through a quantum-confined island. Dashed lines traces the diamond edge for clarity.